# Crystallinity and Crystallographic Texture in Isotactic

# Polypropylene during Deformation and Heating

J. Jia[a,b], D. Raabe [a)]

a. Max–Planck–Institut für Eisenforschung, Max–Planck–Str. 1, 40237 Düsseldorf, Germany

b. Department of Materials, University of Science and Technology Beijing, 100083, Beijing, China

## Abstract

The development of the crystallinity and of the crystallographic orientation of isotactic polypropylene (iPP) during rolling deformation and subsequent heat treatment is studied. The experiments are conducted by using X-ray diffraction with an area detector. The evolution of crystallographic orientation is tracked by calculating the pole figures and by applying a quantitative 3D texture component fit method. The rolling orientation after a true strain of -1.5 mainly consists of the (010)[001], (130)[001], and [001]//RD fiber components (RD: rolling direction). The results reveal that the crystallinity drastically decreases during rolling. We suggest that decrystallization (disaggregation) is a deformation mechanism which takes place as a microscopic alternative to crystallographic intralamellar shear depending on the orientation of the lamellae relative to the imposed deformation tensor. Heat treatment after rolling leads to the recrystallization of amorphous material and to a strong enhancement of the fiber orientation component. The recrystallization orientation is explained in terms of an oriented nucleation mechanism where amorphous material aligns along existing crystalline lamellae blocks which prevailed during the preceding deformation.

# 1. Introduction

The physical and mechanical properties of semi-crystalline thermoplastic polymers depend on the degree of crystallinity [1-3] and on the orientation distribution of the crystalline volume portion [4-6]. Both quantities typically change when the sample is exposed to thermal and mechanical loads. This means that the degree of crystallinity and the orientation distribution are history-dependent properties in such materials.

Various experimental methods have been used to determine the degree of crystallinity of semi-crystalline thermoplastic polymers. These are for instance the measurement of the density and enthalpy of fusion; infrared and Raman spectroscopy; wide-angle X-ray scattering (WAXS); and NMR spectrometry [1, 7]. Methods such as the measurement of the density and enthalpy of fusion via differential scanning calorimetry (DSC) and WAXS are those which are most frequently used for estimating the degree of crystallinity in semi-crystalline thermoplastic polymers. The use of the DSC method for the determination of the crystalline volume portion is reliable for providing accurate values for materials in the *equilibrium* state. However, it is less reliable for materials which are in a strong non-equilibrium state, such as severely strained iPP as in this study. Under such conditions thermally activated relaxation processes render DSC measurements inaccurate [1].

Therefore, the θ-2θ Bragg wide angle X-ray diffraction method has been used in most previous investigations to measure the degree of crystallinity [1, 7-9]. However, the X-ray method may also suffer from drawbacks, namely, when the sample is crystallographically oriented. In such cases the intensity distribution along conventional θ-2θ scans has no quantitative validity except in situations where the specimen has either a totally random or a very strong fiber crystallographic orientation. For the investigation of polymers with more complex crystallographic orientations (which is the rule and not the exception) the measurement of the degree of crystallinity should, therefore, not be conducted by conventional θ-2θ wide angle X-ray scans. Instead the crystallographic orientation of the material must be properly taken into account by an adequate integration procedure which adds up the total reflected intensity over the entire pole sphere.

A number of investigations have been devoted to elucidating the relationship between the evolution of deformation orientations in the crystalline portion of polymers and the underlying microscopic mechanisms. Possible plastic deformation mechanisms in the crystalline phase of semi-crystalline polymers were summarized by Peterlin [10], Bowden and Young [11], Haudin [12], and by Lin and Argon [13]. These authors reported that the crystalline portion of many semi-crystalline polymers, such as PET, HDPE, Nylon-6 and PP, deform mainly by crystallographic shear mechanisms. Although other mechanisms such as twinning and stress-induced martensitic transformation were reported as well, they seem to occur predominantly in highly oriented samples. According to these publications [4,5,11-16], microscopic deformation in the crystalline parts of semi-crystalline polymers occurs in the form of crystallographic shear predominantly on {hk0} planes. This mechanism is often referred to as crystallographic chain slip. Considerations associated with the anisotropy of the interfacial energy suggests that crystalline slip along the (100), (010), and (110) planes should occur preferentially in these materials. Indirect experimental studies have suggested that the two most important chain slip systems are (100)[001] and (010)[001]. The third slip system along the (110) plane, although widely postulated, has not been detected experimentally. It was also reported that transverse slip mechanisms such as, (100)[010] and (010)[100], can occur in addition to the chain slip process.

In this investigation, we aim to exploit specific properties of the area detector for X-ray measurements in terms of the integration and analysis of diffraction diagrams with multiple Bragg cones. An integral calculation method is used to estimate the crystallinity evolution of iPP samples during rolling and subsequent heat treatment based on the integration of wide angle X-ray frames obtained from an area detector. In addition we analyze the evolution of crystallographic orientation during rolling and heat treatment in terms of pole figures and the orientation distribution function derived from them by using the texture component method. We interpret the observed evolution of the crystallinity and crystallographic orientation of iPP as a function of the imposed rolling and heat treatment processes in terms of the underlying microscopic deformation and recrystallization mechanisms.

## 2. Experimental

### 2.1 Material and processing procedure

The study was conducted on a set of commercial semi-crystalline isotactic polypropylene (iPP) sheets with dimensions 100 mm×100 mm×5 mm. The isotactic index amounted to 95% and the density to 0.9 g/cm$^3$. In order to provide homogeneous starting conditions and to reduce the influence of possible microstructures inherited from the preceding manufacturing process all iPP sheets were heat treated with free sample ends at 150°C for 24 hours before rolling. The samples were rolled at room temperature. Each pass exerted a thickness reduction of about 0.1 mm at a constant rolling speed. The total true rolling strain was described by the formula: true strain = ln (h/h$_0$), where h$_0$ is the original thickness of iPP sample and h is the thickness of rolled sample. At various logarithmic (true) strains between -0.4 and -1.5 samples were cut for X-ray analysis. After rolling, iPP sheets were also broadened on the transverse direction. All the X-ray samples were machined from the center part of sheets to minimize the influence of broadening. Samples with a true strain of -1.2 were put in the ovens with temperature of 90$^o$C and 120$^o$C, respectively, at times between 5 to 240 minutes, then were cooled in the air.

### 2.2 X-ray diffraction experiments

The orientation distribution of certain crystallographic planes of the α-iPP crystallites in the deformed and heat treated state were measured by using an X-ray diffractometer set-up with an area detector (Co$_{K\alpha1}$; tube current: 40mA; tube voltage: 40kV) [17]. For covering the entire pole sphere we collected 119 incomplete planar X-ray diffraction frames for each sample at an integration time of 150 seconds per frame. From the normalized and corrected X-ray data the (110), (040), (130) and (-113) pole figures were reconstructed. The sample reference system was orthorhombic owing to the symmetry imposed by the rolling deformation and the crystal symmetry is monoclinic ($\alpha$ modification). The parameters of the unit cell are a=6.65 Å, b=20.80 Å, c=6.5 Å, $\beta$=99.8° [18].

The X-ray diffraction data were also used to calculate the degree of crystallinity of the rolled and heat treated samples. The procedure worked by first integrating each of the 119 X-ray frames into 119 corresponding intensity-2$\theta$ data sets and second by summing these curves into one intensity distribution diagram which presents the integral and orientation-corrected intensity distribution as a function of the 2$\theta$ angle. For the ensuing analysis we adopted a two-phase concept usually applied for polymers where the amorphous contribution to the spectrum which occurs as a broad diffraction band was approximated by a background curve which separates the amorphous from the crystalline portion. The crystalline volume fraction for each specimen was then obtained as the ratio between the area under the crystalline peaks and the total area under the diffraction curve.

## 2.3 Calculation of the orientation distribution function in terms of the texture component method

The spherical distribution of crystals in polycrystalline aggregates is referred to as crystallographic orientation. It is quantified in terms of the orientation distribution function (ODF) which provides the orientation density as a function of the rotation angles. It can be calculated from pole figures by means of series expansion methods or by direct inversion methods [19,20]. In this study we have chosen a third approach, namely, the texture component method [21]. The approach uses sets of spherical Gauss functions with individual height and full width at half maximum as a measure for the strength and scatter of orientation component. The ODF is approximated by a linear superposition of such functions. A detailed description of the method was given in [17,21,22].

## 3. Results and Discussions

### 3.1 Evolution of crystallographic orientation during rolling

Fig. 1 shows the pole figures of some main crystallographic planes of the monoclinic polypropylene crystals as determined for rolled samples at different strains. The pole figures of the (110), (040), and (130) planes illustrate the inclination of the *a* and *b* crystallographic axes with respect to the sample coordinate system. The pole figure of the (-113) planes provides the best possible measure for the orientation of the chain axis because the monoclinic structure does not provide any reflections of the (*00l*) type and the normal to the (-113) plane is only 5.8° misoriented from the direction of the macromolecular chain axis which is equivalent to the crystallographic *c* axis [23].

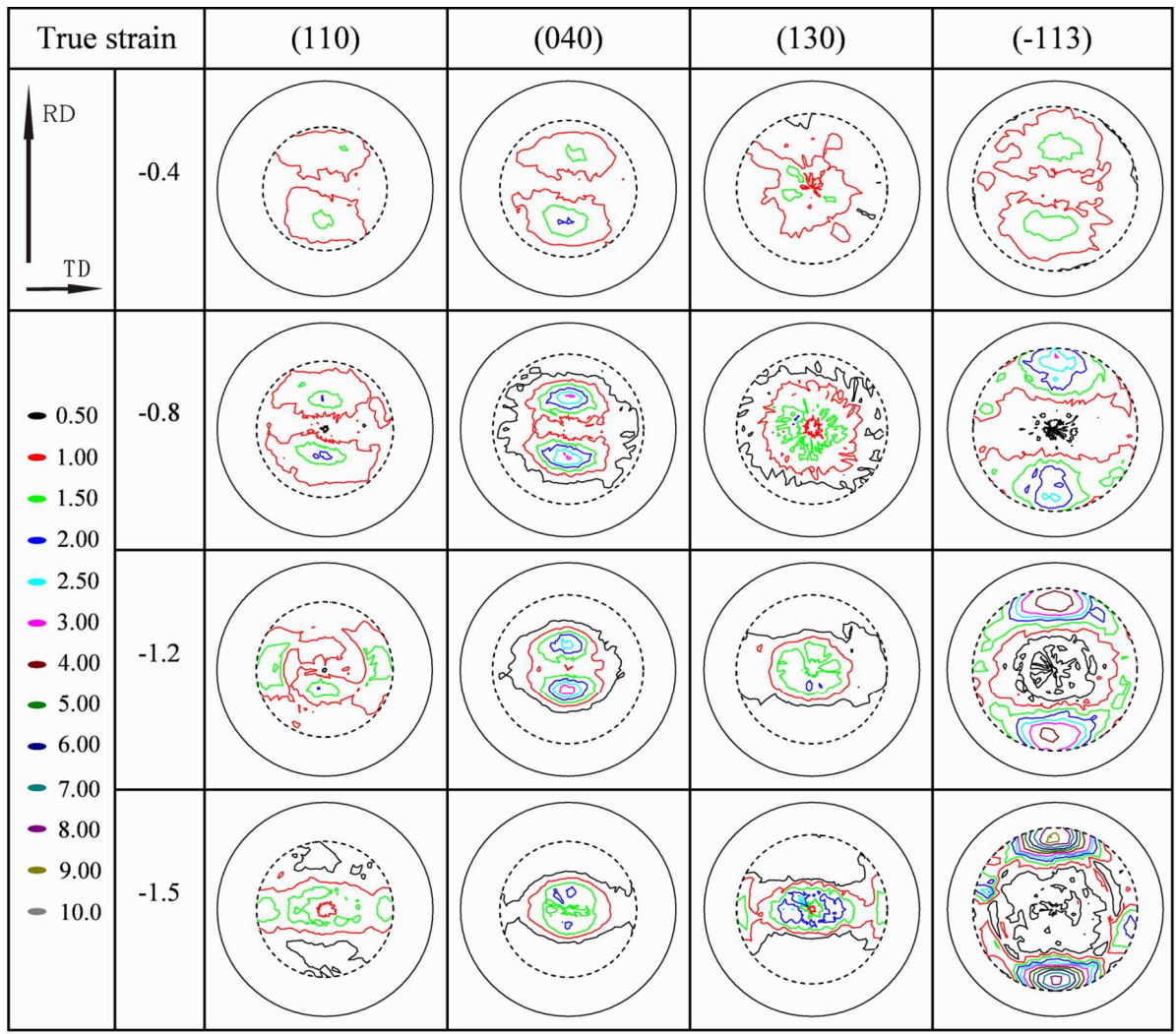

**Fig. 1.** Pole figures of the (110), (040), (130), and (-113) crystallographic planes of the monoclinic α-polypropylene determined for rolled samples at various true (logarithmic) strains. The pole figures are plotted as stereographic projections. RD indicates the rolling direction and TD the transverse direction.

The crystallographic orientation of the material before rolling was practically random. At a true strain of -0.4 a weak crystallographic orientation has formed with a maximum pole density below 2.5 times random. The pole figures for the (110), (040), and (-113) planes have a similar shape at this strain level (Fig. 1, first row), while the (130) projection shows pole clusters around the ND (sheet normal direction). At a true strain of -0.8 the main maxima of the pole figures are more pronounced (Fig. 1, second row). The (040) and (130) planes cluster around the ND, while the (-113) poles gradually reorient towards the RD (rolling direction). As the true strain increases to -1.2 (Fig. 1, third row), all pole figures change gradually except for the (-113) projection. The maxima of the (040) and (130) planes are close to the ND and also begin to cluster around TD (transverse direction), especially the (130) planes. The (110) planes reorient towards the TD and the (-113) planes form a stronger orientation cluster. At a true strain of -1.5 the maxima of the (-113) planes increased along the RD, while the (110), (040), and (130) planes cluster around the TD (Fig. 1, fourth row).

Comparing to the pole figures of iPP film which suffered uniaxial and sequential biaxial drawing [24], rolled samples in this work have the similar orientation of uniaxial drawing iPP film but totally different orientation of sequential biaxial drawing iPP film. That is to say, the effect of broadening can be ignored in this work.

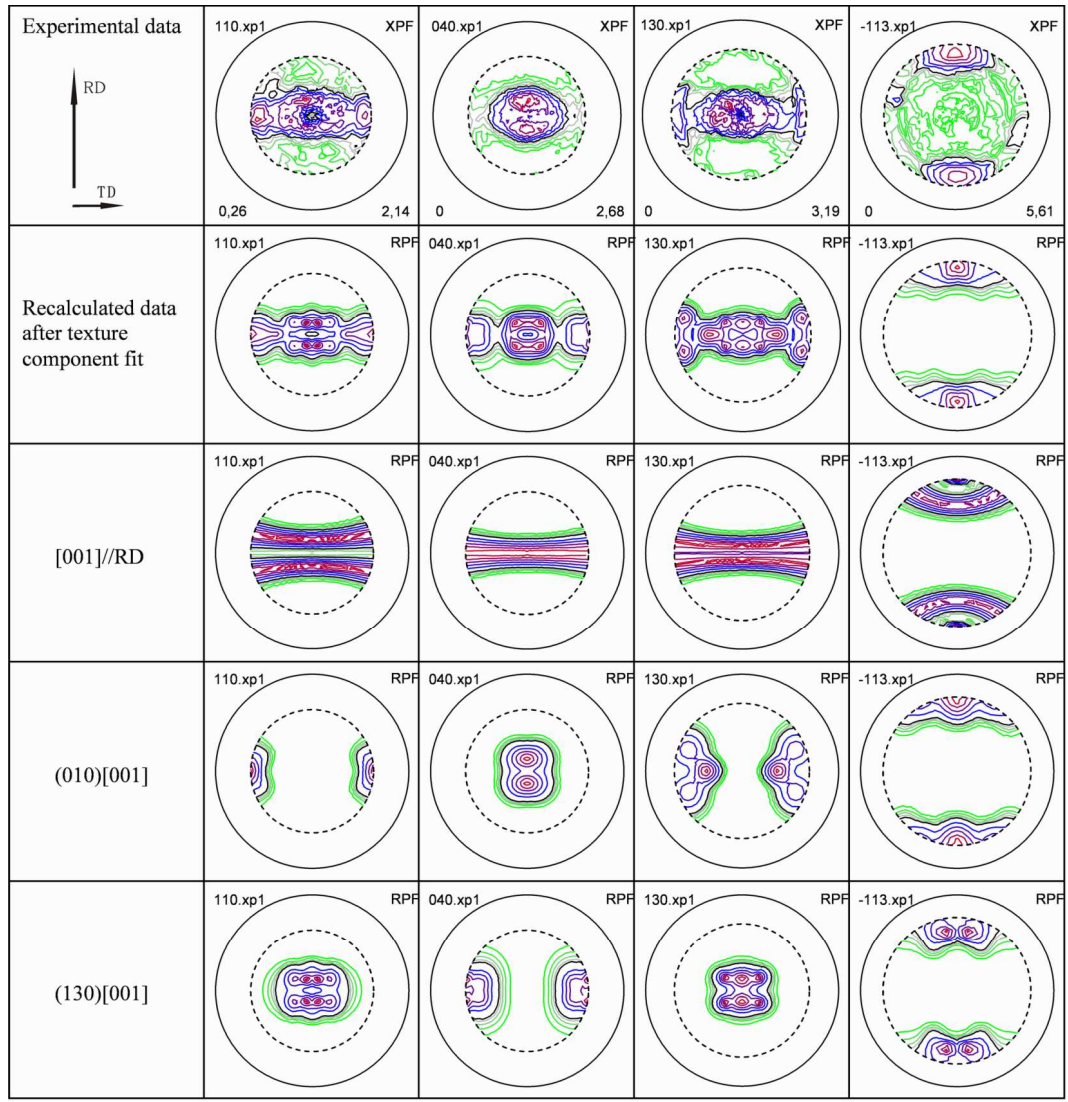

Component #1 ([001]//RD): 12.96 vol.%, spread: 16.27°

Component #2 ((010)[001]): 9.25 vol.%, spread: 15.325°

Component #3 ((130)[001]): 10.84 vol.%, spread: 15.09°

**Fig. 2.** Example of the orientation component fit from experimental pole figures on the basis of the approach of Helming et al. [21] using spherical Gauss-shaped functions. The crystallographic orientation consists of three main orientation components, namely, (010)[001] component, (130)[001] component and a fiber component [001]//RD. The true rolling strain amounts to -1.5.

Fig. 2 shows a typical orientation component fit for a sample rolled to a true strain of -1.5 (Fig. 2, second row) and also provides a list of the main orientation components. The approximation decomposed the crystallographic orientation into three main components, namely, (010)[001], (130)[001] and a fiber orientation component [001]//RD (Fig. 2 third, fourth and fifth rows). Owing to the fact that the crystal symmetry is monoclinic and not cubic the abbreviated use of Miller triples for the orientation components such as (010)[001] must be understood as a preferred crystallographic plane orientation (010) parallel to the surface of the sheet and a preferred crystallographic axis direction [001] parallel to the rolling direction of the sheet. These preferred indices do, however, not necessarily belong to the same crystal.

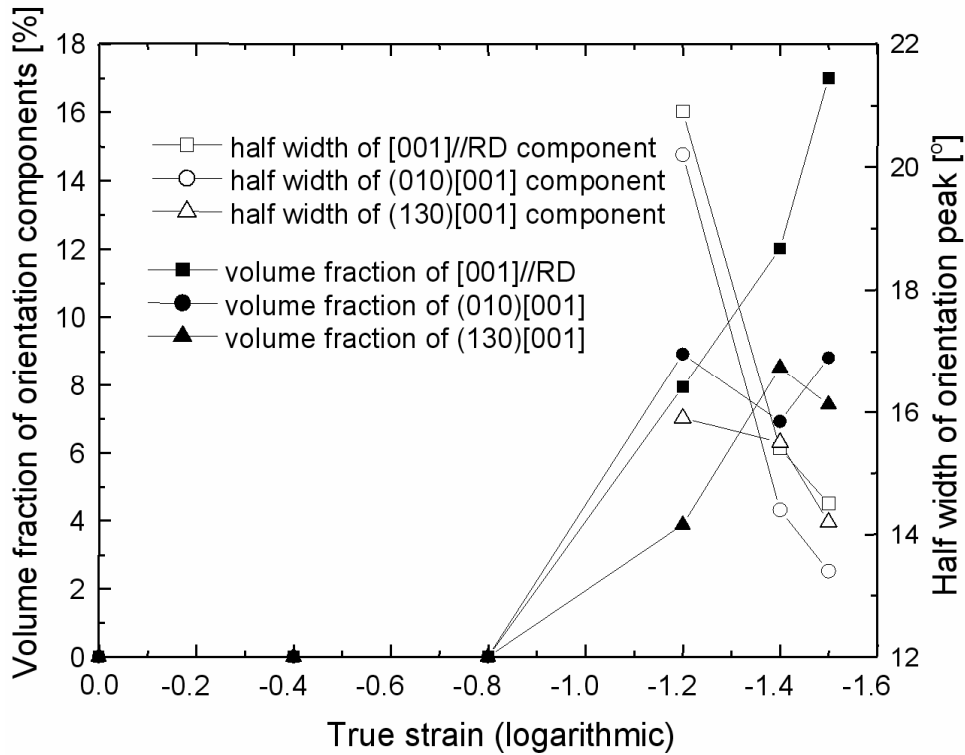

**Fig. 3.** Development of the crystallographic orientation in terms of the volume fractions and spherical half widths of the three main orientation components during rolling. The data were calculated by use of the texture component method from the pole figures shown in Fig. 1.

Fig. 3 provides an overview of the evolution of the crystallographic orientation in terms of the volume fractions and half widths of the three main orientation components during rolling. The data were calculated by use of the texture component method from the pole figures shown in Fig. 1. The diagram reveals that the volume fractions of these orientation components increase with increasing rolling strain. This applies especially to the fiber orientation component. The spherical half widths of the (010)[001] and [001]//RD components become gradually smaller after a strain of -1.2, which indicates an increase in orientation sharpness.

The analysis reveals that the orientation distribution of the crystallites in rolled iPP leads to a multicomponents orientation including as main components (010)[001], (130)[001], and a fiber orientation component [001]//RD. This result suggests that the dominant underlying microscopic deformation mechanism is crystallographic shear along the [001] chain direction. The first orientation component developing in rolled iPP at a low strain of -0.4 is that with a preferred alignment of (010) planes perpendicular to ND while the chain axis distribution tends to orient along RD. The development of such an orientation distribution is probably a result of the activity of crystallographic slip operating on (010) planes along the chain direction [001], i.e. the (010)[001] slip system. This conclusion is consistent with theoretical considerations, which predicted the (010)[001] slip system as the one with the smallest shear resistance [25].

At higher true strains, the crystallographic orientation data show the development of two new orientation components, namely of the (130)[001] component and of the orientation fiber [001]//RD. We, therefore, assume that the originally predominant microscopic shear on the (010)[001] slip system is gradually accompanied by increasing shear activity on the (110)[001] and (100)[001] systems. It is likely that the chain slip systems discussed above are supported by the transverse slip systems which are operating in the same planes as the chain slip systems (e.g. the (010)[100] transverse slip system supporting the (010)[001] chain slip system). This co-operation would explain the observed evolution of crystallographic orientation during rolling deformation.

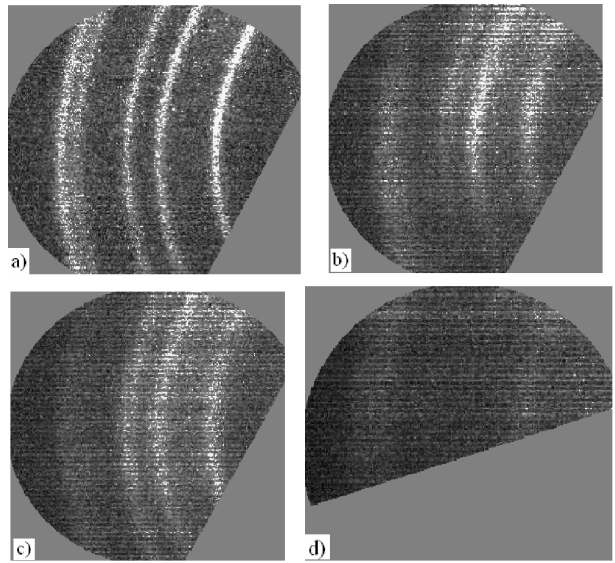

**Fig. 4.** Debye-Scherrer frames taken from the undeformed and from heavily rolled samples (true strain: -1.5) at different rotations. a) undeformed sample, b)-d) rolled sample at different angles, b) Chi=64.6°, Phi=0°, c) Chi=64.6°, Phi=90°, d) Chi=18.2°, Phi=0°.

## 3.2 Evolution of the crystalline volume fraction during rolling

Fig. 4 exemplary shows four X-ray frames obtained from the area detector. The Bragg peaks of the undeformed sample (frame in Fig. 4a) are much sharper than those observed in the rolled sample (true strain: -1.5, frame in Fig. 4b) at the same angular position. For the oriented sample (true strain: -1.5, frames b, c and d) different 119 frames were taken to cover the entire pole sphere. The differences in the patterns reveal how strongly diffraction data may depend on the position of the measurement in the pole sphere. This observation underlines that the measurement of the crystalline volume portion cannot be obtained by conventional $\theta$-$2\theta$ X-ray wide angle line scans but requires instead of taking the crystallographic orientation of the material properly into account as outlined in the preceding chapter. The integrated X-ray data obtained for the undeformed and for the deformed iPP samples at different true strains are shown in Fig. 5.

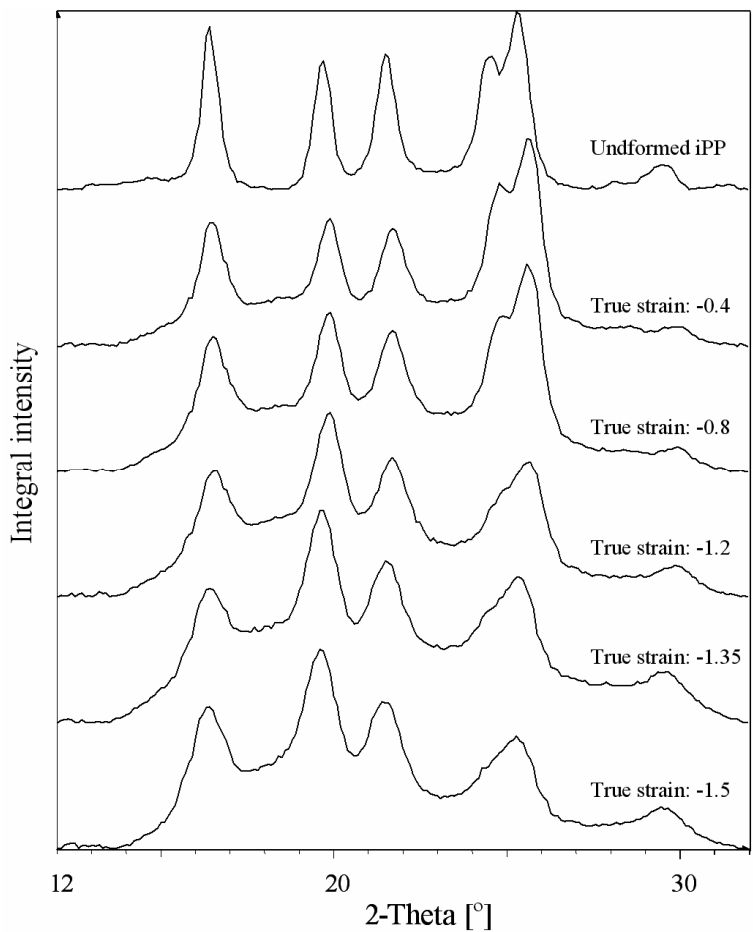

**Fig. 5.** Orientation-corrected integrated wide angle X-ray diffraction curves as a function of the true rolling strain. Integration is conducted over all 119 X-ray diffraction frames which were measured for each sample state.

The figure reveals that the curves become gradually smoother during rolling, i.e. the Bragg peaks are less sharp and pronounced relative to the diffraction band which is created by the amorphous phase at large strains. The reason for this argumentation is that the loss in the crystalline volume fraction occurs by the gradual disintegration of the lattice cells. This means that crystals change their lattice parameter gradually until the cells are in the end completely disintegrated. This may lead to the observed Bragg broadening effect owing to the spread in the lattice parameter. Also the gradual disintegration of the crystalline phase may induce internal stresses which also modify the lattice parameter entailing similar Bragg broadening effects. On the other hand it must be underlined that this decrease in the Bragg scatter relative to the background scatter is *not* due to a change in crystallographic orientation since the data are integrated over the entire pole sphere comprising the signals of crystals of all possible orientations, i.e. the integral method applied in this work ensures that the *overall* crystallinity of the material is being evaluated.

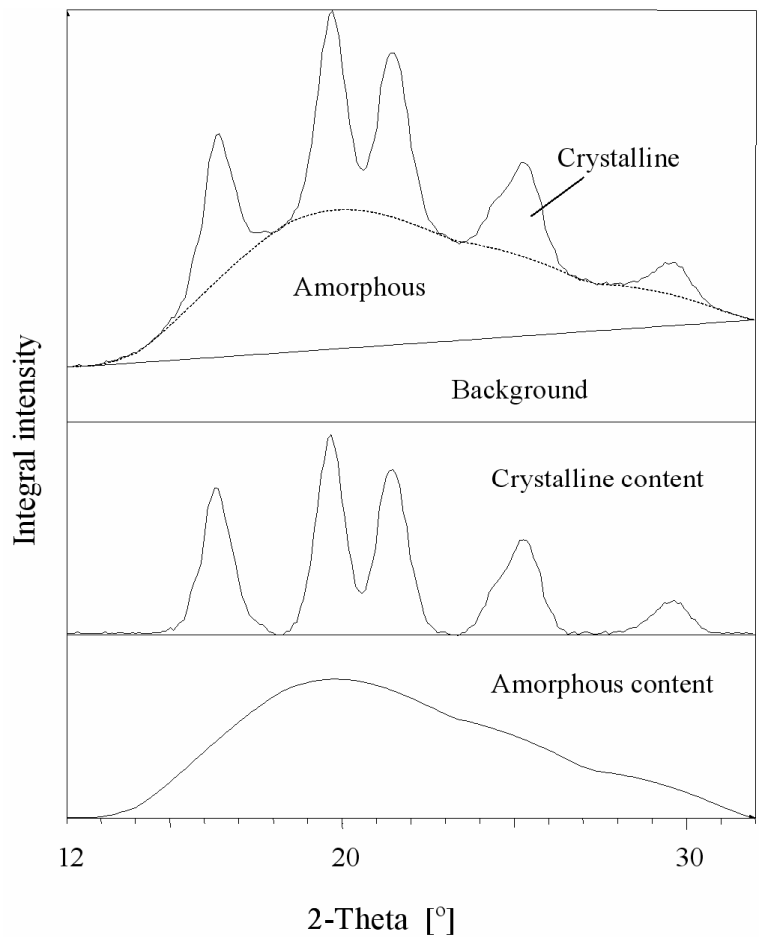

**Fig. 6.** Typical fitting procedure for integral WAXS results obtained for a rolled iPP sample (true strain: -1.2).

Fig. 6 is an example of a typical fitting procedure as it is used in this work for the separation of the crystalline peaks from the amorphous halo for an iPP sample with a true strain of -1.2. The dotted line separates the crystalline component from the amorphous phase. The separations of crystalline part from the amorphous part were made by the EVA program with Hermans-Weidinger method [26]. The crystallinity for each specimen was then obtained from the ratio between the area under the crystalline peaks and the total area under the diffraction curve.

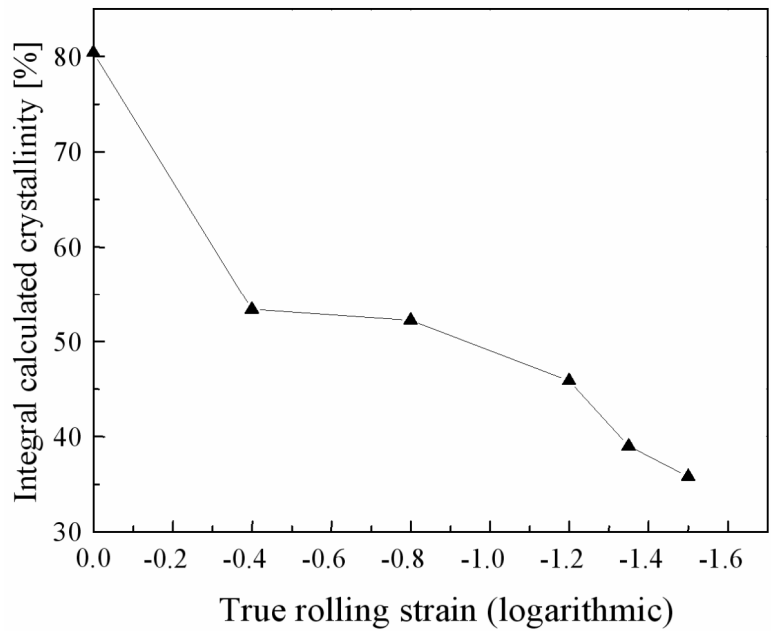

**Fig. 7.** Evolution of the volume content of the crystalline phase as a function of the magnitude of the true rolling strain as calculated from the integrated X-ray diffraction data.

Fig. 7 shows the result for the relative change in the crystallinity as a function of deformation in terms of the true rolling strain. The crystallinity of the iPP samples is substantially reduced with increasing true strain, varying from an initial value of 80.1vol.% for the undeformed sample to the final value of 35.8vol.% for the deformed sample at the true strain of -1.5. Owing to the fact that the data have been obtained by integration over the entire pole sphere the influence of the crystallographic orientation is fully accounted for. The degree in crystallinity drops particularly strong between a strain of 0.0 and -0.4. Since we do not have any additional data points between these two strain levels it is conceivable that the strong drop occurs not right away at the initial stages of deformation but later between 0.3 and 0.4. However, this is speculative at the current stage. The strong drop at the initial stages of the deformation will hence be investigated by additional density measurements in an ensuing study.

The strain-induced decrystallization of crystalline material in semi-crystalline polymers has also been referred to as disaggregation [17]. Similar observations and related conclusions were during the last years published by the group of Strobl [27-31] on the basis of data obtained from uniaxial tensile testing. In these investigations the authors suggest that the process of tensile deformation consists of several regimes which are characterized by the activation of different types of deformation mechanisms. At small loads they assume that intralamellar slipping of crystalline blocks is the prevalent mechanism of deformation. At larger strains above the yield strain they observe a stress-induced crystalline block disaggregation-recrystallization process. In a recent work Men et al. [27] suggested that the strain at this transition point between the two regimes is related to the interplay between the amorphous entanglement density and the mechanical stability of the crystal blocks. The authors furnished this two-stage deformation model by careful experimental evidence obtained from true stress-strain curves.

In previous investigations about rolled PET (polyethylene terephthalate) [17,32], the present authors suggested that deformation-induced decrystallization must be regarded as a regular microscopic deformation mode in semi-crystalline PET. Decrystallization (disaggregation) is a deformation mechanism which takes place as an alternative to crystallographic intralamellar shear depending on the orientation of the lamellae.

In this work, we make use of this approach to explain the decrystallization mechanism during rolling. We assume that the easiest mode of microscopic deformation for a crystalline lamella is the crystallographic shear on the (010)[001] slip systems. At higher stresses we anticipate that the slip systems (100)[001] and (110)[001] are additionally activated. When stressed along a certain direction within the five-dimensional stress space which does *not* favor the activation of any of these shear systems, the respective crystalline portion undergoes decrystallization as an alternative mode of deformation. In other words if the local stress tensor in an iPP crystal portion reaches the yield surface in any direction other than that entailing compatible crystallographic slips the lamella becomes decrystallized, i.e. it starts to disintegrate as a *compatible* mode of deformation. Compatibility in context refers to the fact that the displacement gradient fields of neighboring material portions match so that no overlap or hole is created.

In macro-mechanical terms the decrystallization, therefore, corresponds to the violation of the von-Mises criterion which requires five independent internal microscopic shear modes for an imposed deformation tensor with its five independent components. In this context one should note that the competition between a crystallographic shear mechanism and a decrystallization mechanism is not only a matter of stress but of the shape of the yield surface. The anisotropic flow stress is a 6-dimensional convex stress surface which is limited by the disintegration stress in cases where the Schmid factor of the existing crystallographic shear systems is zero. The authors also suggest that both mechanisms may even occur at the same time, however, differently in different crystalline lamellae, according to their respective crystalline orientation factors. This means that while some crystals start to disintegrate already at rather small strains according to their unfavorable Schmid factors for regular crystallographic slip, others may remain mechanically stable also up to higher strains, but they reorient according to the plastic spin resulting from their respective active slip or block shear system.

## 3.3 Recrystallization and evolution of crystallographic orientation during heat treatment

The measurement of the change in crystallinity of the rolled iPP during heat treatment is based on the integration of the 119 X-ray diffraction frames over the entire pole sphere as explained in section 2.2.

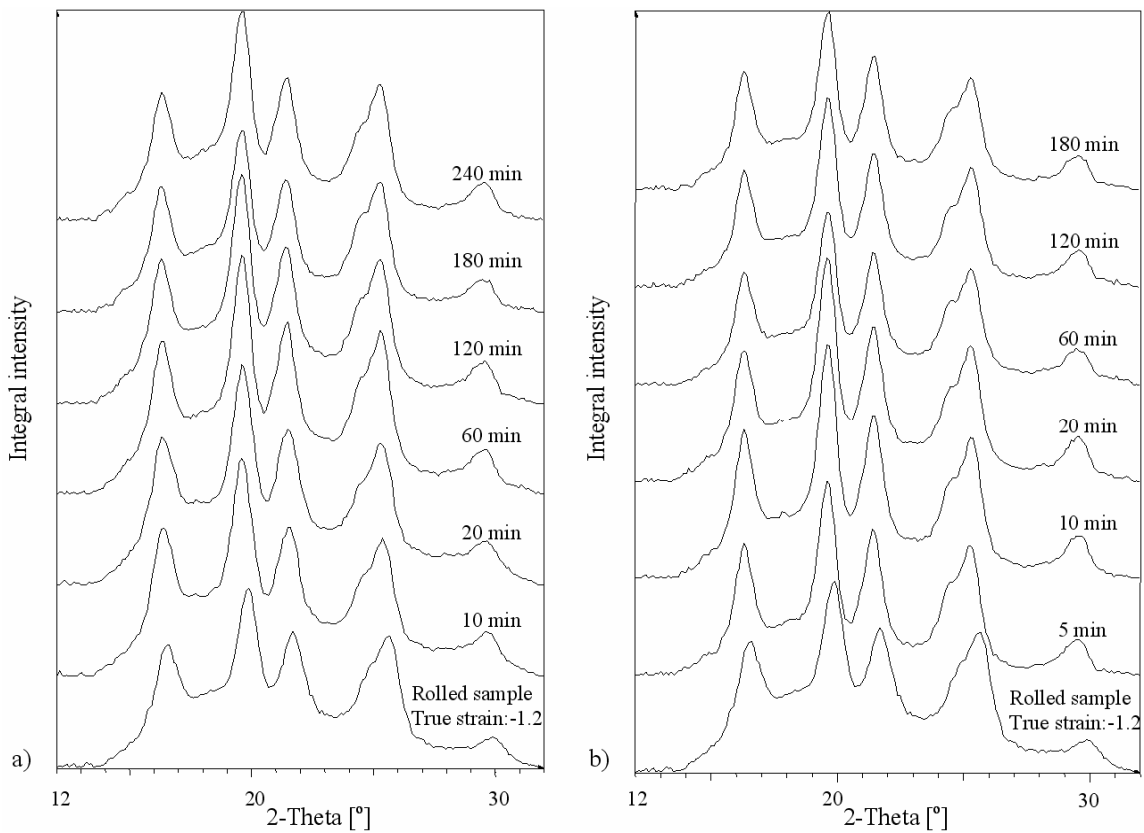

**Fig. 8.** Integrated wide angle X-ray scans of rolled iPP samples (true strain: -1.2) as a function of the heat treatment time. a) 90°C, b) 120°C.

Fig. 8 shows the integrated wide angle X-ray scans of a rolled iPP sample (true strain: -1.2) and of a rolled plus heat treated sample for different annealing times at 90°C (Fig. 8a) and 120°C (Fig. 8b), respectively. At the beginning of the heat treatment of the rolled specimens, the Bragg peaks become gradually sharper and the amorphous diffuse portion becomes quickly smaller, especially at 120°C, indicating a rapid increase in the crystalline volume fraction. At the same time, a minor shift of the (*hkl*) reflections to smaller values of the 2θ angles suggests that the distance between the (*hkl*) planes becomes somewhat larger. For heat treatment times beyond 20 minutes almost no further structural changes can be observed. The data show that both phenomena are much faster and more pronounced at 120°C than at 90°C.

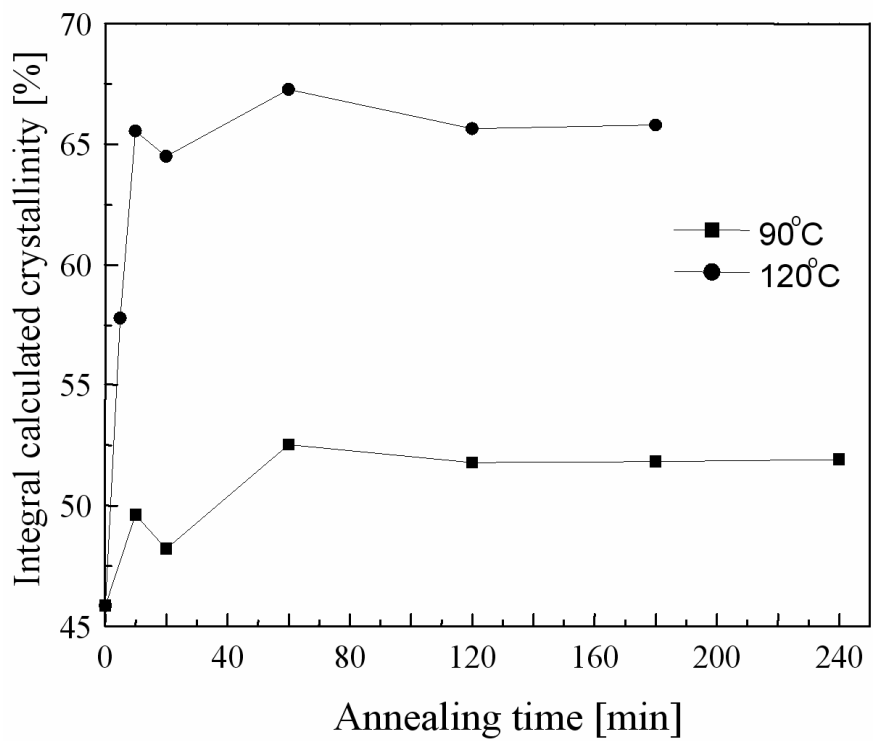

**Fig. 9.** Changes in the crystallinity with the heat treatment time at two different temperatures. The data were obtained from integrated X-ray diffraction data.

Fig. 9 shows the results for the corresponding relative changes in crystallinity of the rolled specimens as a function of the heat treatment time at 90°C and 120°C, respectively. The crystalline volume fractions of the iPP samples drastically increase particularly at the beginning of the heat treatment, varying from an initial value of 45.8vol.% for the rolled sample to values of 49.6vol.% and 65.6vol.% for the heat treated samples at 90°C and 120°C, respectively, after only 10 minutes. For heat treatment times beyond 10 minutes, the crystalline volume fractions remain almost constant at 120°C and slightly increase further in the samples treated at 90°C. This result indicates that the thermally activated *recrystallization* of the decrystallized material occurs at a high rate at the beginning of the heat treatment.

Fig. 10 shows the evolution of crystallographic orientation of rolled iPP (true strain: -1.2, rolled at room temperature) during a subsequent heat treatment at 90°C and 120°C, respectively, in terms of pole figures. When comparing the pole figures of the rolled sample (Fig. 1, third row) with those taken from the heat treated samples, it can be observed that no main new orientation components emerge during the heat treatment.

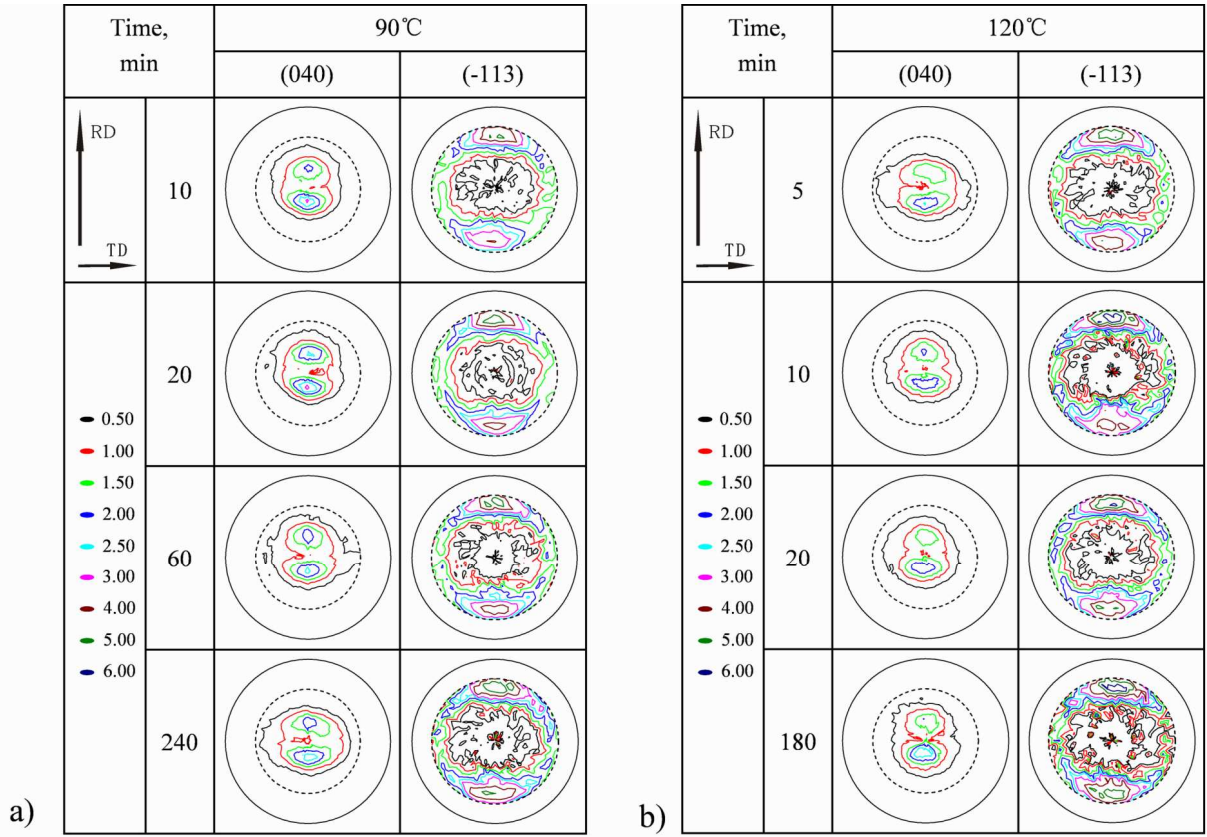

**Fig. 10.** Presentation of the evolution of crystallographic orientation in terms of pole figures of rolled (true strain: -1.2) and subsequently heat treated iPP specimens. a) 90℃, b) 120℃.

At the beginning of the thermally induced recrystallization process, the intensities of (040) pole figures slightly decrease. But the (-113) pole figures show a somewhat higher intensity than the rolled sample after the initial 10 minutes of the heat treatment, which are 4.75, 5.8 and 6.48 for rolled sample, 60℃ and 100℃ annealing samples, separately. For longer heat treatment times the pole figures do not reveal main further changes. This observation matches the data shown in Fig. 9 which revealed that the crystallinity reaches a constant level after 10 minutes at 120℃. This means that the increase in crystallinity can be mainly attributed to an increase in intensity of the (-113) pole figures.

Fig. 11 shows the evolution of crystallographic orientation in terms of quantitative orientation data obtained from the texture component method. The diagrams show the changes in the volume fractions of the three main orientation components in the course of the heat treatment. At the beginning of the heat treatment, all curves reveal strong changes in their relative volume fractions. The data between 20 minutes and 60 minutes show a transition regime in the evolution of crystallographic orientation. For the (010)[001] and (130)[001] orientation components (Fig. 11a), the final volume fractions are slightly lower than for the rolled sample which is reflected by the slightly lower intensity in the (040) pole figures (Fig. 10). For the [001]//RD component (Fig. 11b), the final volume fractions of the heat treated samples are much larger than those observed for the rolled samples. The shape of the curve for the evolution of the fiber component (Fig. 11b) correlates well with the curve observed for the evolution of the crystalline volume fraction during heat treatment (Fig. 9).

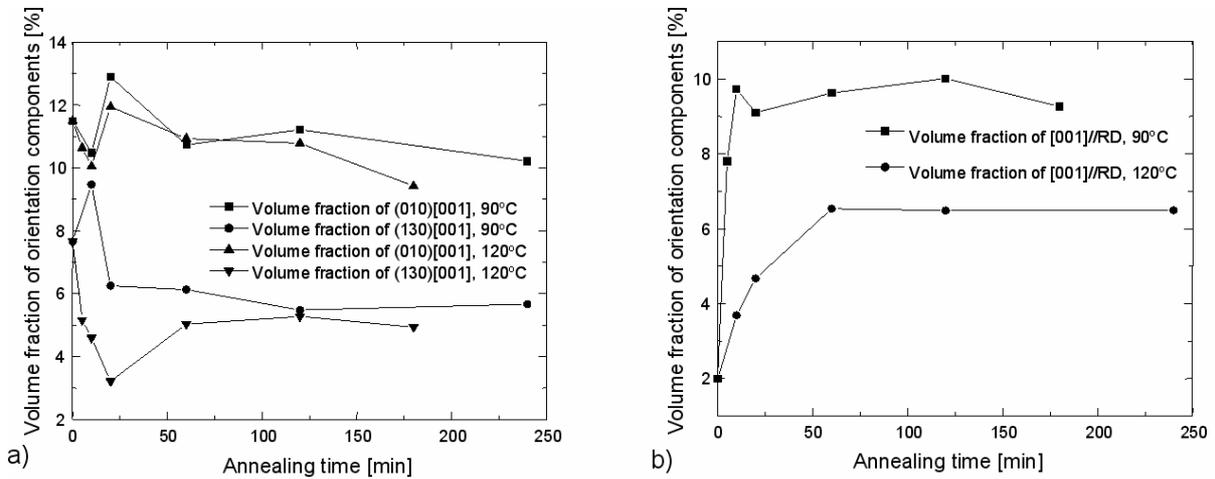

**Fig. 11.** Changes in the volume fractions of the three main orientation components during the heat treatment of the rolled samples. a) (010)[001] component, (130)[001] component, b) fiber component [001]//RD.

It can be observed that both, the pole figures (Fig. 10) and the quantitative orientation component data (Fig. 11) can be consistently explained in terms of the results obtained for the crystallinity (Fig. 9). Heat treatment (90°C and 120°C) of the sample rolled to a true strain of -1.2 leads first, to the recrystallization of the amorphous portions and second, to an enhancement of the original fiber orientation component during recrystallization. The results further show that the strong recrystallization process does obviously not lead to entirely new crystalline orientations, but merely to a sharpening of the preceding rolling orientation. This applies in particular to the fiber orientation component. We can, therefore, draw the conclusion that recrystallization works in the present case by an oriented nucleation.

The data also show that recrystallization essentially happens during the first 10 minutes of the heat treatment. Owing to the pronounced temperature dependence of the recrystallization kinetics we suggest that there is a critical annealing time for the onset of recrystallization. Also we observe different saturation levels for the crystallinity of iPP which are characteristic for the different heat treatment temperatures. In this work, the critical isothermal annealing time for iPP amounts to about 10 minutes and the saturation levels for the recovery of the maximum possible crystallinity are about 51.5% and 65% at 90°C and 120°C, respectively.

## 4. Conclusions

We conducted an experimental crystallographic orientation study in conjunction with an integral calculation of the crystallinity on rolled and heat treated iPP samples. The data were discussed in terms of several possible crystallographic shear modes, deformation-induced decrystallization processes, and thermally activated recrystallization [33]. The main conclusions are:

- The final crystallographic orientation of the cold rolled iPP sheets at a true strain of -1.5 is characterized by three main orientation components, namely, (010)[001], (130)[001] and [001]//RD. These orientation components support the assumption that the principal deformation mechanisms active during deformation were crystallographic slip on the (010)[001], (100)[001] and (110)[001] systems, all propagating along the chain direction. These slip systems are probably supported by other slip systems acting in the same planes in a direction transverse to the chain direction.

- It was observed that the crystallinity of the samples continuously drops during rolling from about 80% (undeformed) to about 35% at a true strain of -1.5. The process was referred to as decrystallization or disaggregation. We explain this phenomenon as a deformation mechanism which may take place as an alternative to crystallographic slip depending on the crystallographic orientation of a lamella owing to the local violation of the von-Mises criterion.

- Heat treatment (90°C and 120°C) of rolled samples leads to the recrystallization of amorphous material and to an enhancement of the original deformation orientation. This observation was explained in terms of an oriented nucleation or respectively relaxation mechanisms where amorphous material aligns along existing crystalline lamellae or fragmented lamellae blocks.